\documentstyle[prl,aps]{revtex}
\begin{document}
\draft

\twocolumn[\hsize\textwidth\columnwidth\hsize\csname@twocolumnfalse\endcsname

\title{Transverse depinning and melting of a moving vortex lattice in driven 
periodic Josephson junction arrays}

\author{Ver\'{o}nica I. Marconi and Daniel Dom\'{\i}nguez}
\address{Centro At\'{o}mico Bariloche, 8400 S. C. de Bariloche,
Rio Negro, Argentina}

\date{\today}
\maketitle
\begin{abstract}
We study the effect of thermal fluctuations
in a vortex lattice driven  in the periodic pinning of a 
Josephson junction array. 
The phase diagram current ($I$) vs. temperature ($T$) is studied.
Above the critical current $I_c(T)$ 
we find a moving vortex lattice (MVL) with
anisotropic Bragg peaks.
For large currents $I\gg I_c(T)$, there is  a melting transition of the
MVL at $T_M(I)$. When applying a
small transverse current to the MVL, there is 
negligible dissipation at low $T$.
We find  an onset of transverse vortex motion at 
a transverse depinning  temperature $T_{tr}(I)<T_M(I)$.
\end{abstract}


]                

\narrowtext

The interplay between the periodicity of 
vortex lattices (VL) and periodic pinning potentials 
in superconductors raises many interesting questions both
in equilibrium \cite{wire,jja,martinoli,dot,hole,franz}
and in driven systems  \cite{nori}.
Experimentally, 
periodic pinning has been realized in artificially fabricated systems
like: 
superconducting wire networks \cite{wire}, 
Josephson junction arrays \cite{jja}, 
thickness modulated superconducting films \cite{martinoli}, 
magnetic dot arrays \cite{dot}
and submicron hole lattices \cite{hole} in superconductors.  
commensurability effects in the  ground state
vortex configurations lead to enhanced critical currents and resistance
minima for ``fractional''  and for ``matching'' 
(i.e. commensurate) vortex densities where the VL is strongly pinned.
Under the effect of thermal fluctuations, 
it is possible to have a depinning phase transition of these
commensurate ground states at a temperature $T_p$ 
and a later melting transition of the VL at a temperature $T_M$
\cite{franz}. 
For high vortex densities
(i.e. strongly interacting VL) both transitions coincide,
$T_p=T_M$, while for low vortex densities
both transitions are different with $T_p<T_M$.
Out of equilibrium, many recent studies have concentrated in the 
related problem of the driven VL in the presence of random 
pinning \cite{KV,GLD,bmr,pardo,simu,dgb}. 
The nature of the fastly moving vortex structure for large driving forces 
has been under active discussion lately \cite{KV,GLD,bmr} motivating
both experimental \cite{pardo} and numerical \cite{simu,dgb,mio} studies.
In particular, the interesting concept of {\it transverse} critical current 
has been introduced in Ref.~\cite{GLD}. 
After applying a current in the direction perpendicular to the drive, a finite
transverse critical current $I_{c,tr}$ may exist, at least at $T=0$ \cite{GLD}.
For $T>0$,  there is a very small but finite transverse
linear response \cite{bmr}, with a possible
sharp non-linear voltage increase at an ``effective'' 
$I_{c,tr}$ \cite{GLD,bmr}.
For periodic pinning, the physics of the driven VL has
been  studied numerically only at $T=0$, 
where a complex variety of dynamic phases  
has been reported \cite{nori}. In this case, it is clear that a finite
$I_{c,tr}$ will exist due to commensurability effects \cite{bmr},
and it has been obtained in \cite{nori} for $T=0$.

In this Letter we study 
the effect of {\it thermal fluctuations}
in a driven VL in a periodic pinning potential. 
In our case, the periodic pinning is provided by  a 
two dimensional Josephson junction array (JJA) \cite{jja,dyna,nos}.
We obtain a phase diagram  
as a function of the driving current ($I$) and temperature ($T$),
which is shown in Fig.~1.
For low currents, we find that
the depinning and melting transitions are separated 
with $T_p(I)<T_M(I)$. More interestingly,
for large currents we find an analogous sequence of transitions
but for the transverse response of a fastly moving VL.
We find that there is a novel  {\it transverse depinning} temperature
$T_{tr}$ below the melting transition of the moving VL,
$T_{tr}(I)<T_M(I)$.

The current flowing in  the junction between two superconducting islands 
in a JJA is 
modeled as the sum of the Josephson supercurrent and the normal current 
\cite{dyna,nos}:
\begin{equation}
I_{\mu}({\bf n})= I_0 \sin\theta_{\mu}({\bf n}) + 
                  \frac{\Phi_0}{2\pi c R_N} 
		  \frac{\partial \theta_{\mu}({\bf n})}{\partial t}
		  +\eta_{\mu}({\bf n},t)
\end{equation}
where $I_0$ is the critical current of the junction between
the sites ${\bf n}$ and ${\bf n}+{\bf \mu}$ in a square lattice 
[${\bf n}=(n_x,n_y)$, ${\bf \mu}={\bf \hat x}, {\bf \hat y}$], 
$R_N$ is the normal state resistance
and $\theta_{\mu}({\bf n})=\theta({\bf n}+{\bf \mu})-\theta({\bf
n})-A_{\mu}({\bf n})=\Delta_\mu\theta({\bf n})-A_{\mu}({\bf n})$ is the
gauge invariant phase difference with $A_{\mu}({\bf n})=\frac{2\pi}{\Phi_0}
\int_{{\bf n}a}^{({\bf n}+{\bf\mu})a}{\bf A}\cdot d{\bf l}$.
The thermal noise fluctuations $\eta_{\mu}$ have correlations
$\langle  \eta_{\mu}({\bf n},t)\eta_{\mu'}({\bf n'},t')\rangle=
\frac{2kT}{R_N}\delta_{\mu,\mu'}\delta_{{\bf n},{\bf n'}}\delta(t-t')$.
In the presence of an external magnetic field $H$ we have
$\Delta_{\mu}\times A_{\mu}({\bf n})= A_x({\bf n})-A_x({\bf n}+{\bf y})+ 
A_y({\bf n}+{\bf x})-A_y({\bf n})=2\pi f$, $f=H a^2/\Phi_0$ and 
$a$ is the array lattice spacing.
We take periodic boundary  conditions (p.b.c) in both directions in the presence
of an external current $I_{ext}$ in the $y$-direction in arrays with 
$L\times L$ junctions \cite{mio}.
The vector potential is taken as
$A_{\mu}({\bf n},t)=A_{\mu}^0({\bf n})-\alpha_{\mu}(t)$ where 
in the Landau gauge $A^0_x({\bf n})=-2\pi f n_y$, $A^0_y({\bf n})=0$
and $\alpha_{\mu}(t)$ will allow for total voltage fluctuations. 
With this gauge the p.b.c. for the phases are: 
$\theta(n_x+L,n_y)=\theta(n_x,n_y)$ and
$\theta(n_x,n_y+L)=\theta(n_x,n_y)-2\pi f Ln_x$. 
The condition  of a current flowing in the $y$- direction:
$\sum_{\bf n} I_{\mu}({\bf n})=I_{ext}L^2\delta_{\mu,y}$
determines the dynamics of $\alpha_\mu(t)$ \cite{mio}.
After considering conservation of current, 
$\Delta_\mu\cdot I_{\mu}({\bf n})=\sum_{\mu} I_{\mu}({\bf n})-
I_{\mu}({\bf n}-{\bf \mu})=0$, we obtain:
\begin{eqnarray}
\Delta_{\mu}^2\frac{\partial\theta({\bf n})}{\partial t}&=&-\Delta_{\mu}\cdot
[S_{\mu}({\bf n})+\eta_{\mu}({\bf n},t)]\\
\frac{\partial\alpha_{\mu}}{\partial t}&=&I_{ext}\delta_{\mu,y}
-\frac{1}{L^2}\sum_{\bf n} [S_{\mu}({\bf n})+\eta_{\mu}({\bf n},t)]
\end{eqnarray}
where $S_{\mu}({\bf n})=\sin[\Delta_\mu\theta({\bf n})-A_{\mu}^0({\bf n})-
\alpha_{\mu}]$, we have normalized currents by $I_0$, time
by $\tau_J=2\pi cR_{N}I_0/\Phi_0$, temperature
by $I_0\Phi_0/2\pi k_B$, and the discrete laplacian is
$\Delta^2_\mu\theta({\bf n})=\theta({\bf n}+{\bf\hat x})
+\theta({\bf n}-{\bf\hat x})+\theta({\bf n}+{\bf\hat y})
+\theta({\bf n}-{\bf\hat y})-4\theta({\bf n})$.

The Langevin dynamical equations (2-3) are solved with a second order 
Runge-Kutta-Helfand-Greenside
algorithm with time step $\Delta t=0.1\tau_J$ and integration time $10000\tau_J$
after a transient of $5000\tau_J$. The discrete laplacian is inverted with
a fast Fourier + tridiagonalization algorithm as in \cite{nos}.
We study the following properties:
(i) {\it Superconducting coherence}: we calculate the helicity modulus
in the direction transverse to the current 
$\Upsilon_x=\frac{1}{L^2}\langle\sum_{{\bf n}}\cos\theta_x({\bf n})\rangle
-\frac{1}{T}\frac{1}{L^4}\{\langle [\sum_{{\bf n}}\sin\theta_x({\bf
n})]^2\rangle
- \langle [\sum_{{\bf n}}\sin\theta_x({\bf n})]\rangle^2\}$.
[In order to calculate the helicity modulus along $x$, we enforce
strict periodicity in $\theta$ by fixing $\alpha_x(t)=0$]. 
(ii){\it Transport}: we calculate
the time average of the total voltage  $V=\langle v_y(t)\rangle=
\langle d\alpha_y(t)/dt\rangle$ (voltages are normalized
by $R_ {N}I_0$).
(iii) {\it Vortex structure}:
we obtain the vorticity
at the plaquette ${\bf \tilde n}$ (associated to the site ${\bf n}$)
as $b({\bf \tilde n})=-\Delta_\mu\times{\rm nint}[\theta_\mu({\bf n})/2\pi]$
with ${\rm nint}[x]$ the nearest integer of $x$. We calculate the average
vortex structure factor as $S({\bf k})=\langle|\frac{1}{L^2}\sum_{\bf \tilde n}
b({\bf \tilde n})\exp(i{\bf k}\cdot{\bf \tilde n})|^2\rangle$.

We study  JJA with a magnetic field corresponding to $f=1/25$ and system
sizes of $L\times L$ junctions, with $L=50,100$. 
The ground state vortex
configuration for $f=1/25$ is a tilted 
square-like vortex lattice (VL) commensurate
with the underlying periodic pinning potential of the square JJA
(see \cite{teitel25}).
The structure factor $S({\bf k})$ has correspondingly delta-like
Bragg peaks.
For this value of $f$ we find an equilibrium
phase transition at $T_c\approx0.050\pm0.003$, which
corresponds to a simultaneous VL depinning 
(corresponding to the onset of resistivity and 
vanishing of helicity modulus) 
and VL melting (corresponding to the vanishing of Bragg peaks);
i.e. $T_c=T_p=T_M$.

First, we have calculated the current-voltage (IV) characteristics
for different temperatures. At $T=0$ there is a critical current of $I_c(0)=
0.114\pm0.002$, which corresponds to the single vortex depinning current
in square JJA \cite{jjpin}. Above $I_c(0)$ 
there is an almost linear increase of
voltage until $I\approx 1$ where there is a sharp rise of $V$ because
all the junctions become normal. Similar behavior has been reported
for $T=0$ IV curves for low values of $f$ \cite{yu}. We restrict our 
analysis for
currents $I<0.4$, where the collective behavior of the VL is  the dominant
physics.  
For temperatures $T<T_c$ we see that there is a sharp rise
in voltage for the apparent critical current $I_c(T)$, which decreases with
$T$, vanishing at $T_c$. In Fig.1 we plot the $I_c(T)$ line 
obtained with a voltage criterion of $V<10^{-4}$. 
For currents below $I_c(T)$ there is a pinned vortex lattice
(PVL) which is the same as
the $T=0$ ground state, with delta-like Bragg peaks.
On the other hand, for currents $I>I_c(T)$ there is a {\it moving vortex
lattice} (MVL), which has {\it anisotropic} Bragg peaks in
the structure factor $S({\bf k})$ as shown in Fig.~2(a).
There are two features in  the anisotropy of $S({\bf k})$ :
(i) The {\it height} of the peaks decreases in the direction of vortex motion
(i.e. perpendicular to the current drive).
(ii) The {\it width} of the peaks increases in the direction perpendicular
to vortex motion. This means that thermal broadening is less effective
in the direction of motion.
We have also studied the behavior of the Bragg peaks of the PVL and the
MVL for three different lattice sizes ($L=50,100,150$) and different values
of $I$ and $T$. In the PVL the peak height is independent
of the system size, as expected for a pinned lattice. On the other hand,
for the MVL, the peak height decreases  with system size,
with a power law behavior $S({\bf G})\sim L^{-\eta_G(I,T)}$
and $0<\eta_G(I,T)<2$. In Fig.~2(b) we show a finite size analysis for
two Bragg peaks $S(G_1)$ and $S(G_2)$ at a particular
point in $(I,T)$ with a power law fit.
This is the expected behavior for 
a floating solid in two dimensions \cite{franz}.
In general we see
that $\eta_G$ increases with $T$ for a given current. 
When $\eta_G(I,T)>2$ the MVL melts into a liquid.
The anisotropic structure of the MVL of Fig.~2(a) 
is similar to the behavior predicted for a moving Bragg glass \cite{GLD}. 
However in our case there is no random
pinning, but periodic pinning. The only source of dynamic randomness are
thermal fluctuations.

We now study in more detail the  different transitions by fixing a
given value of the current $I$ and slowly changing temperature $T$
with small increases of $\Delta T = 0.0005$.
In this way we obtain the phase diagram shown in Fig.~1. We have also
cross-checked these results with the IV curves at fixed $T$.
There are two cases of interest: (i) low currents $I<I_c(0)$ and 
(ii) large currents $I>I_c(0)$.

(i){\it Low currents}.
In Fig.~3 we show the behavior for $I=0.03 < I_c(0)$. At low temperatures
the voltage is almost zero since the VL is pinned. When increasing $T$
there is a sharp rise of the voltage (of two orders of magnitude) at
a depinning temperature $T_p(I)$, as shown in Fig.3(a). At this temperature
the VL starts to move since the driving current is higher than the 
critical current. Therefore, this corresponds to a transition from
a pinned VL to a moving VL. We find that the $I_c(T)$ line obtained
from the IV curves at fixed $T$ coincides with the $T_p(I)$ line
obtained from the $V-T$ curves.  
We have also calculated at the same time 
the structure factor $S({\bf k},T)$ and 
the transverse helicity modulus $\Upsilon_x(T)$. 
In Fig.~3(b) we show the behavior of two Bragg peaks $S(G_1)$  and
$S(G_2)$. For $T<T_p(I)$ we see that $S(G_1)=S(G_2)$ since there is a
pinned VL with isotropic structure factor. Above $T_p(I)$ 
we find that  $S(G_1)\not=S(G_2)$.
This shows the fact that there is a MVL with anisotropic 
Bragg peaks. These peaks vanish at a higher temperature 
$T_M(I)$ in a continuous
and smooth transition, corresponding to a melting of the MVL.
Above $T_M(I)$ all Bragg peaks vanish and there is a vortex liquid for
$T>T_M$.
In the inset of Fig.~3(b) we show $S(G_1)$ for two different sizes
$L=50,100$ 
[we find similar finite size effects  for $S(G_2)$].  
We see that for $T<T_p(I)$ the $S(G_1)$ is size independent
since the VL is pinned, while for $T_p(I)<T<T_M(I)$ there is 
a power-law size dependence as expected for  a floating solid,
see Fig.~2(b). The temperature $T_M(I)$ where
$S(G_1)$ vanishes is size independent.
On the other hand, the $\Upsilon_x(T)$ has a significant drop
at $T_p(I)$, however it remains
finite but with large fluctuations in the MVL phase, $T_p<T<T_M$. This 
suggests that in the MVL there is  superconducting coherence in the 
direction transverse to the current.

(ii) {\it Large currents}.
When the VL is driven with a large current $I>I_c(0)$ there is a moving
VL  with anisotropic Bragg peaks even at $T=0$. 
In Fig.~4 we show our results for $I=0.16>I_c(0)$.
The structure factor is always anisotropic as can be seen in Fig.~4(b)
where $S(G_1)\not=S(G_2)$. 
We find that the Bragg peaks vanish at a temperature $T_M(I)$, which
is size independent, as shown in the inset of Fig.~4(b) for $S(G_1)$
[similar behavior is found for $S(G_2)$].
We have investigated the possibility of a 
transverse critical current $I_{c,tr}$ \cite{GLD,bmr}. 
At $T=0$ a finite $I_{c,tr}$ is expected due to commensurability
effects  \cite{bmr,nori}. 
We have applied
a  transverse current $I_{tr}=I_x$ in the $x$-direction
(in addition to the applied bias, $I=I_y=0.16$) and we have calculated
the transverse voltage response $V_{tr}$.
We find that for finite low temperatures $V_{tr}$ 
is negligible small within our numerical accuracy, until
there is a sharp increase at an ``effective''
 transverse critical current  $I_{c,tr}(T)$. We find that
$I_{c,tr}(T)$  tends to vanish at a temperature $T_{tr}$.
An interesting way of studying this phenomenon is to apply a small current
$I_{tr}$ and vary $T$. 
 In Fig.~4(a) we study the onset of the {\it transverse  depinning transition}:
we apply a small current  $I_{tr}=0.01$, and we show
the transverse resistive response $V_{tr}/I_{tr}$ vs $T$.
We see that $V_{tr}$ is vanishingly small at low $T$ and it rises at $T_{tr}$.
This transition temperature 
is clearly {\it below} $T_M$ as we can see in Fig.~4.
We have obtained $T_{tr}(I)$ for two driving currents $I>I_c(0)$ as we show
in Fig.~1. It seems reasonable that this transverse depinning line will exist
all along this region of the phase diagram. For the intermediate temperatures
$T_{tr}<T<T_M$ we find that there is always an ordered vortex array
but the  orientation and structure of the MVL
depends on the initial conditions.
Moreover, finite size analysis shows that for $T<T_{tr}$ the
MVL structure factor is weakly size dependent with $\eta_G(T) \sim 0.05$,
while after  $T_{tr}$, the exponent $\eta_G$ has a steep increase
to values $\eta_G(T) \sim 0.5$.
The helicity modulus is shown in Fig.~4(c). 
We see that for $T<T_{tr}$ the MVL has transverse superconducting coherence 
with a well-defined $\Upsilon_x$. 
On the other hand,  $\Upsilon_x$ 
shows strong fluctuations in the region $T_{tr}<T<T_M$.

In conclusion, we have studied the current-temperature phase diagram
of a vortex lattice driven in a periodic Josephson junction array.
We find that for low currents the ``longitudinal'' depinning transition of the
VL and the later melting of the moving VL are different with
$T_M(I)>T_p(I)$. 
For large currents we find a very
analogous behavior for the ``transverse'' response of a fastly moving VL
(compare Fig.~3 with Fig.~4). 
In this case it is possible to define
a transverse depinning transition at a temperature $T_{tr}$, and
a later melting transition of the moving VL at $T_M(I)$.
This transverse depinning transition could be easily studied
in controlled experiments in Josephson junction arrays with
transport measurements.

We acknowledge Hernan Pastoriza for useful discussions and
Fundaci\'{o}n Antorchas  and Conicet (Argentina) 
for financial support.

\begin{figure}

\caption{ $I-T$ Phase diagram for
$f=1/25$. $T_M(I)$ line is obtained from $\Upsilon_x$ vs. $T$
curves $(\Box)$ and from $S({\bf G})$ vs. $T$ curves $(\bullet)$.
$T_p(I)$ line is
obtained from IV curves $(\star)$, from $S({\bf G})$ vs. $T$
curves $(\bullet)$ and from $\langle V_y\rangle$ vs. $T$ curves $(\triangle)$.
$T_{tr}(I)$ curve is obtained from $\langle V_{tr}\rangle$ 
vs. $T$ curves $(\circ)$.
Temperature is normalized by $I_0\Phi_0/2\pi k_B$.}
\end{figure}

\begin{figure}
\caption{(a) Intensity plot of the structure factor $S({\bf G})$ for
a moving vortex lattice at  $I=0.06$, $T=0.02$.
(b) Finite size analysis and power law fit of $S({\bf G})\sim L^{-\eta_G}$;
we obtain $\eta_{G_1}=0.25\pm0.06$ ($\ast$) and
$\eta_{G_2}=0.14\pm0.05$ ($\triangle$).}
\end{figure}

\begin{figure}
\caption{For $I<I_c(0)$, $I=0.03$:
(a) Dissipation $\langle V_y\rangle/I$ vs. $T$.
(b) Structure factor at two lattice vectors, $S({\bf G_1})$ $(\star)$ and
$S({\bf G_2})$ $(\diamond)$ vs. $T$.
Inset: Size effect in $S({\bf G})$.
(c) Helicity modulus $\Upsilon_x$ vs. $T$.
}
\end{figure}

\begin{figure}
\caption{For $I>I_c(0)$ , $I=0.16$:
(a) Transverse dissipation $\langle V_{tr}\rangle /I_{tr}$ vs. $T$.
(b) $S({\bf G_1})$ $(\star)$ and $S({\bf G_2})$ $(\diamond)$ vs. $T$.
Inset: Size effect in $S({\bf G})$.
(c) $\Upsilon_x$ vs. $T$.
}
\end{figure}

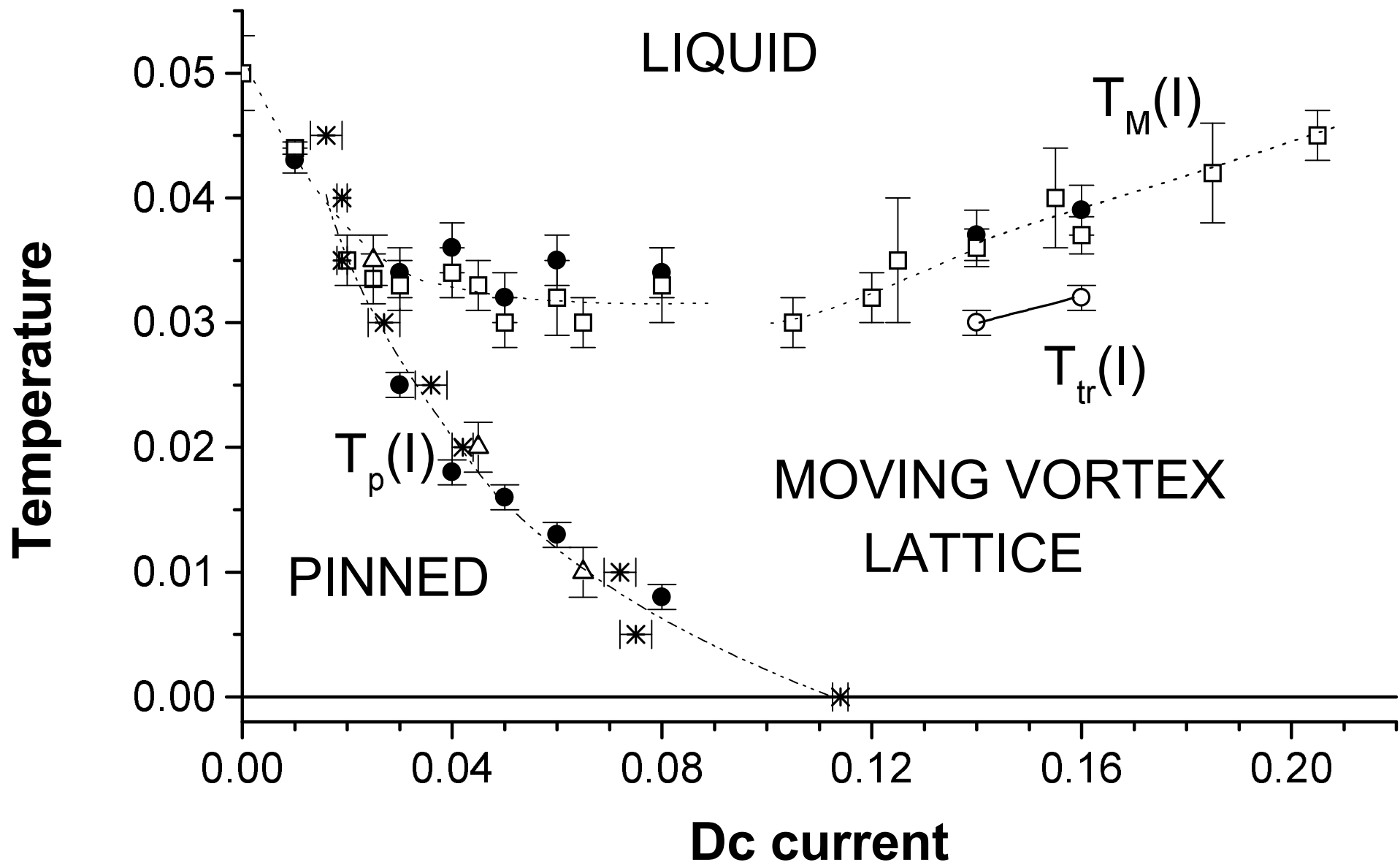
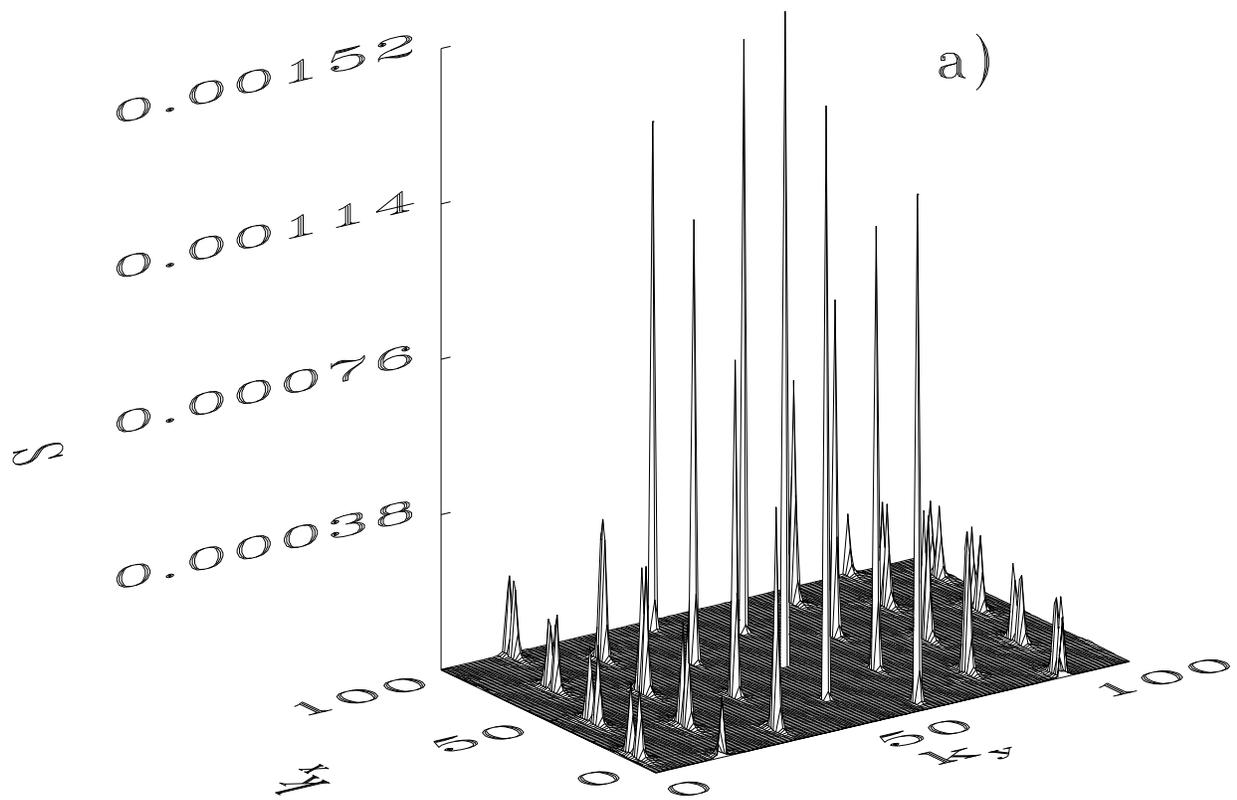
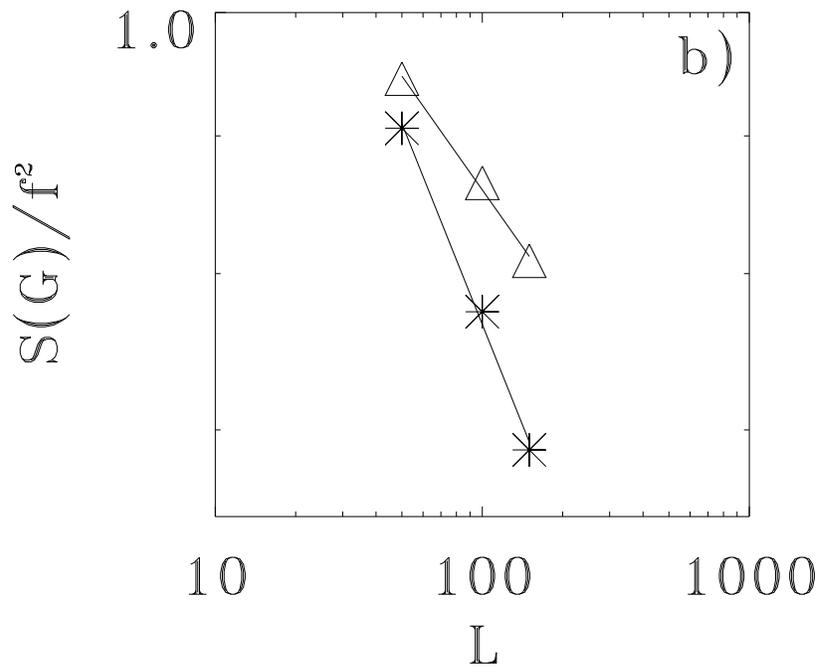
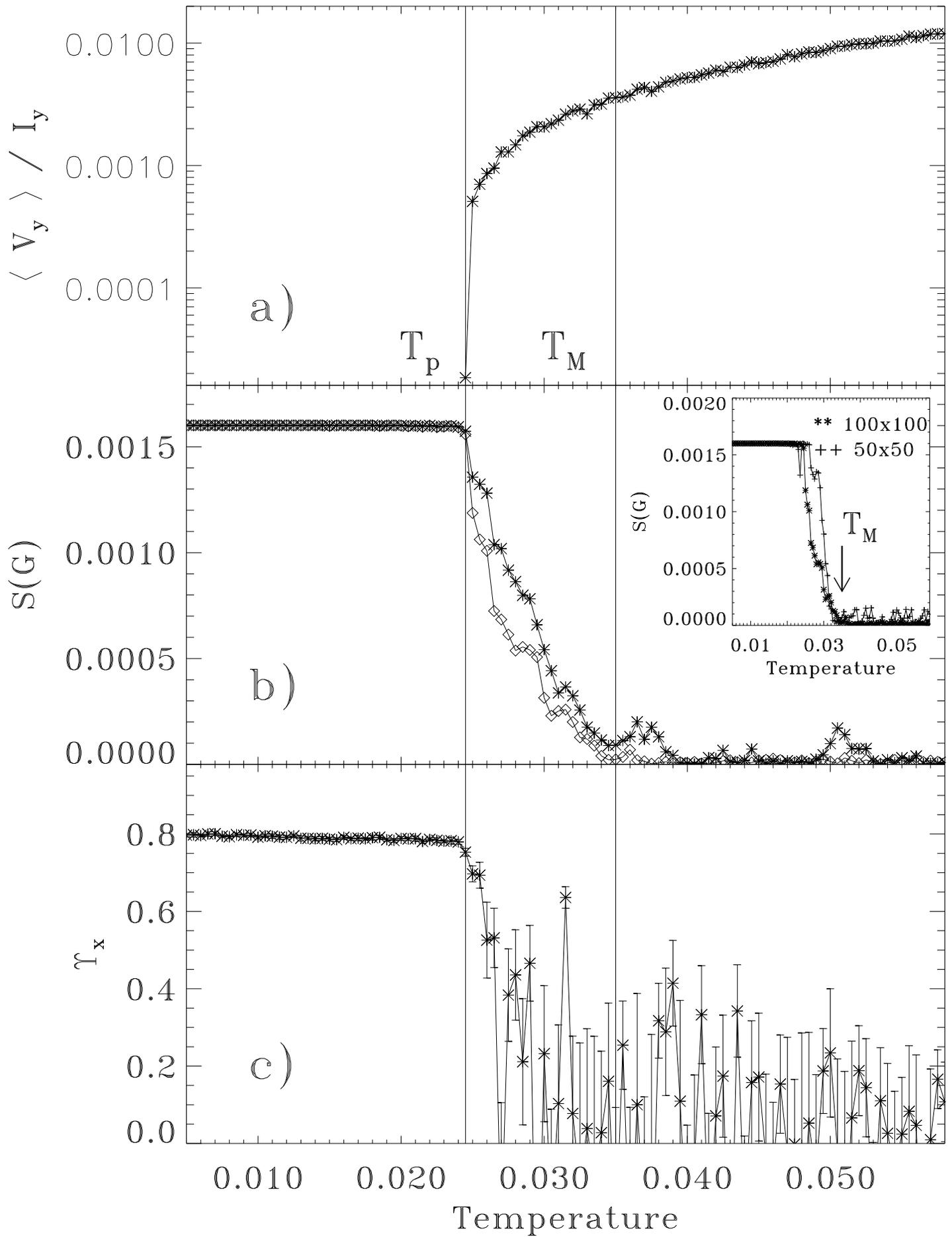
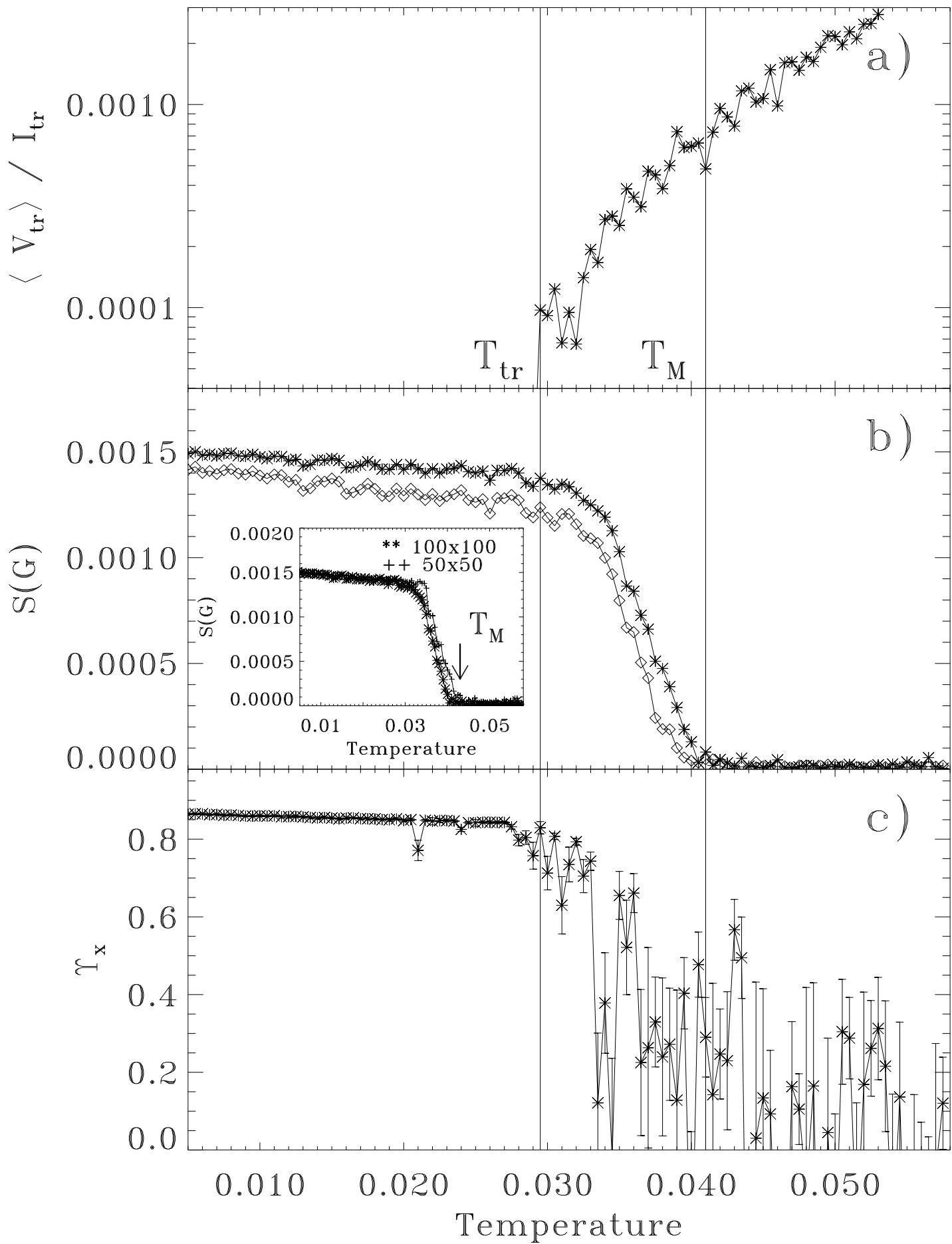
\end{document}